# **Directed Nano-antennas for Laser Fusion** - (vs 3)

Zsuzsanna Márton[1], Imene Benabdelghani[2], Márk Aladi[2], Judit Budai[3], Aldo Bonasera[4,5], Attila Bonyár[6], Mária Csete[3], Tibor Gilinger[3], Martin Greve[5], Jan-Petter Hansen[7], Gergely Hegedűs[2], Ádám Inger[2], Miklos Kedves[2], Károly Osvay[3,8], István Papp[2,9], Péter Rácz[2], András Szenes[3], Ágnes Szokol[2], Dávid Vass[3], Parvin Varmazyar[8], Miklós Veres[2], Konstantin Zsukovszki[2], Tamás S. Bíró[2], Norbert Kroó[2], Laszlo P. Csernai[2,7,10,11] (for the FUSENOW and NAPLIFE Collaborations)

[1] ELI ALPS, ELI-HU Non-Profit Ltd., Szeged, Hungary
[2] HUN-REN Wigner Research Centre for Physics, Budapest, Hungary
[3] Department of Optics and Quantum Electronics, University of Szeged, Szeged, Hungary
[4] Cyclotron Institute, Texas A&M University, College Station, 77843 TX, United States
[5] Laboratori Nazionali del Sud, INFN, via Santa Sofia, 62, Catania, 95123, Italy
[6] Budapest University of Technology and Economics, Budapest, Hungary
[7] Department of Physics and Technology, University of Bergen, Norway
[8] National Laser-Initiated Transmutation Laboratory, University of Szeged, Szeged, Hungary
[9] King's College London, London, United Kingdom
[10] Csernai Consult Bergen, Bergen, Norway
[11] Frankfurt Institute for Advanced Studies, Frankfurt am Main, Germany

**Abstract**:

Why do we use nano-antennas for fusion? In three sentences: The present laser induced fusion plans use extreme mechanical shock compression to get one hotspot and then ignition. Still fusion burning spreads slower than expansion, and mechanical instabilities may also develop. With nano-antennas in radiation dominated systems, simultaneous ignition can be achieved in the whole target volume and there is no time left for mechanical instabilities. Ignition is achieved with protons accelerated in the direction of the nanoantennas that are orthogonal to the direction of laser irradiation.

Present laser fusion methods are based on extreme and slow mechanical compression with an ablator surface on the fuel target pellet to increase compression and eliminate penetration of laser electromagnetic energy into the target. This arises from a mistaken assumption, [1] that the detonation normal 4-vector should have vanishing time-like component, and this assumption eliminates the possibility to rapid or even simultaneous, radiation dominated detonations, (which are well known in the burning (*or hadronization*) of Quark Gluon Plasma).

## **Introduction**

Laser induced fusion has ***two basic obstacles***: One is the RT instability, and faster expansion than the spread of fusion burning. The second e.g., at LNL NIF, that laser beams are thermalized in a “hohlraum”, creating a thermal equilibrium electromagnetic radiation at high temperature. Linearly polarized, monochromatic laser beam is just the same type of electromagnetic radiation as the radio or TV broadcast, but at different wavelengths. While the TV broadcast wavelength is of the order of a meter (m) the visible or laser light is much shorter, ~100 nanometer (nm), i.e. seven orders of magnitude shorter!

Energy transfer can be in two forms (i) "*mechanical*" or water, wind, electrical energy and (ii) thermal energy or "*heat*" energy. Mechanical energy can be transferred to other forms of *mechanical* energy as well as to *heat* with nearly 100% efficiency, while heat can be transferred to mechanical energy less effectively, typically with 30-40%, so called Carnot efficiency. The rest becomes waste heat. Thus, if one has mechanical energy and needs mechanical or electric energy, thermalization in the energy transition leads to considerable losses, thus we must avoid it. Linearly polarized, monochromatic laser beam is "mechanical" energy.

The laser beam is monochromatic radiation of transverse linear polarization. Such kinds of electromagnetic energy can be transferred to large distances with little loss, in coaxial cables, wave conductors and in outer space with resonant parabolic antennas and narrow beams. Thus, for laser induced fusion energy, one should avoid thermalization as much as possible during the process. This is one of the aims of the NAPLIFE project.

The two *basic obstacles* are more essential, and their **solution is a unique** feature of the NAPLIFE project. As we see it, today all other projects are aiming for fusion ignited in a hot spot and then spreading the fusion burning wave through the whole target volume. This scenario arises from the Taub's 1948 description of *relativistic detonation* waves [1], which predicted that these burning waves could spread at most with a shock wave speed, which is just somewhat faster than a mechanical sound wave. Consequently, this leaves time to develop mechanical instabilities and rapid expansion from extreme compression and pressure. However, a decade later [2] in 1987 it turned out that Taub's original derivation was incomplete and ***simultaneous detonations*** (or detonations across a time-like hypersurface) are possible.

This was then verified experimentally in ultra-relativistic heavy ion collisions, where the Quark Gluon Plasma (QGP) was burning (hadronizing) at constant *proper* time. In the field of fusion research this is mostly not known.

For laser induced fusion to achieve this type of detonation was not simple or easy [3].
The possibility opened by using nanotechnology, which allowed the regulation of confinement and amplification of laser light in the *fusion fuel target* [4].
This is done by embedding ***resonant nanorod antennas*** or phased array antennas in required density in the fusion fuel target. There are different resonant antennas. In solar panels for the solar (thermal) radiation spherical core-shell antennas are used, which are absorbing all polarization Sunlight from all directions in a relatively broad resonance frequency band ($\lambda$ ~ 400-800 nm), which covert the Sun's thermal radiation frequency peaking at T~5700K or at $\lambda$ ~ 470 nm, to heat, and we can boil water with such antennas in water.

As the **laser beam is not thermal**, it has linear polarization orthogonal to the beam, and one precise frequency and wavelength (e.g. $\lambda$ ~800nm). This is adequate for *energy transfer with minimal loss*. This requires different antennas [5]. The simplest is a thin **half wavelength dipole** (e.g. $\lambda$ /2~400 nm) antenna in vacuum. Still as the antenna is embedded in the fusion fuel (UDMS polymer) target the effective wavelength, $\lambda_{eff}$, is shorter than in vacuum, and depends on the refractive index of the fusion fuel target ($\lambda_{eff}$ can be approximated by considering the antenna as a segment of an infinite insulator-metal-insulator (IMI), supporting a squeezed plasmonic mode of wavelength determined by the diameter, the gold dielectric properties and the embedding medium index e.g. n=1.5). Furthermore, to realize a thin antenna in nm size length, is technologically not possible, so we will have to manufacture a “thick” antenna, which decreases the resonant length of the antenna. Thus, we end up with a nano-antenna of length ($\lambda_{eff}$/2) x diameter (e.g. 85 x 25 nm) and it will result in a somewhat wider absorption frequency band also.

The orientation of these antennas should be parallel to the polarization vector (E-field oscillation direction) of the laser beam. (However, even if the antennas are not precisely oriented, these resonant antennas increase the antenna gain significantly.) These antennas are not only increasing the absorption of light into heat but in their enhanced near-field directly **accelerate the protons** of ionized Hydrogen of the fusion target parallel to the nanoantennas. Thus, we get accelerated, non-thermal protons parallel to the laser beam’s polarization. This way we can achieve a non-thermal configuration at the stage before the nuclear fusion reaction without thermalization loses!

**Modeling of Angular Proton Distribution**

In the experimental setup the bottom of target chamber is horizontal in the $[x,z]$ plane. The incoming laser beam, the polarization of the beam (The ***E*** -field) vector, the two, forward (FWD) and backward (BWD) Thompson Parabola (TP) detectors, the average momentum of the outgoing particles in the plume, the normal vector of the flat target and the direction of the “horizontal” nanorod antennas in the target are all in this horizontal plane. The direction of the “vertical” nanorod antennas are in the orthogonal, $y$-direction. Let us define the directions in the $[x,z]$ plane so that the $x$-direction is (0º/180º), the incoming laser beam comes from 45º and its polarization is at the angles 135º/225º

We model the angular distribution of the directed nanorod dipole antenna using the EPOCH PIC kinetic model, just as in publications [6, 7]. Let us assume that the laser beam arrives in the horizontal direction at 45º and it is polarized in orthogonal direction to the beam. In the “horizontal” target the nanorod antennas are directed in the horizontal $x$-direction at (0º/180º), i.e. the polarization of the laser beam and the direction of the nanorod antennas are 45º degree from one another in the same horizontal plane. We show in the horizontal $[x,z]$-plane the angular distribution of the proton number and proton energy. In contrast, in the same figures, we show also in the same modeling the configuration where the direction of the nanorod antennas is vertical thus orthogonal to the polarization of the laser beam.

According to these expectations we conclude that directed nanoantennas (or even more nanoantenna phased arrays) lead to increased proton acceleration in the direction of the nanoantennas at 135º/225º. It is different to most laser induced fusion experiments, which follow the TNSA or Picher-Catcher configurations. In our experimental set up, the acceleration is actually to the direction of the nanorod antennas. As we had no TP detectors at 45º or 225º degrees, we could not test this in the present setup directly. Still the experimental results as well as the EPOCH simulations confirm that the directed proton acceleration is in the direction of the directed nanorod antennas,

As we aim for two-sided irradiation to achieve simultaneous ignition of the target, this is of basic importance for the NAPLIFE project. In this case the laser irradiation must be orthogonal to the flat target, and the acceleration by the directed nanorods will be orthogonal to the direction of laser irradiation.

**Target manufacturing with directed nanoantennas**

The directed Au nanorod structures were prepared in the ELI-ALPS Ultrafast Nanoscience Laboratory by electron beam lithography (EBL) in a Raith eLine Plus EBL system. The EBL process parameters are the following: (1) substrate: UV-quality optical quartz glass (10 mm x 10 mm x 1 mm); (2) resist: AR-P 672.03 PMMA [8]; (3) spin-coating: at 2500 rpm, then heating on hot plate at 150°C for 3 minutes resulting in a ~150 nm PMMA layer; (4) e-beam writing: at 25 kV, dose 180 mC/cm$^2$, Au nanorods with 102x30x30 nm nominal sizes, arranged in a rectangular array with 500 nm period in both dimensions; (5) development in AR 600-55; (6) Au thermal evaporation, 30 nm; (7) lift-off in acetone. The excess Au and the resist are removed during the lift-off, and fields of Au nanorod arrays are left on the surface.

Next, another 170 nm PMMA layer was spin coated on top of the nanostructured sample (AR-P 672.03, @1900 rpm), so the medium between the Au nanorods is PMMA. The final stratigraphy of the sample is depicted in Fig. 1.

As in the present experiment the laser beam irradiation was not orthogonal but 45º to the target the best resonance length of a thick target is not obvious to estimate, especially as the nanorod antenna is not embedded into a homogeneous material.

.

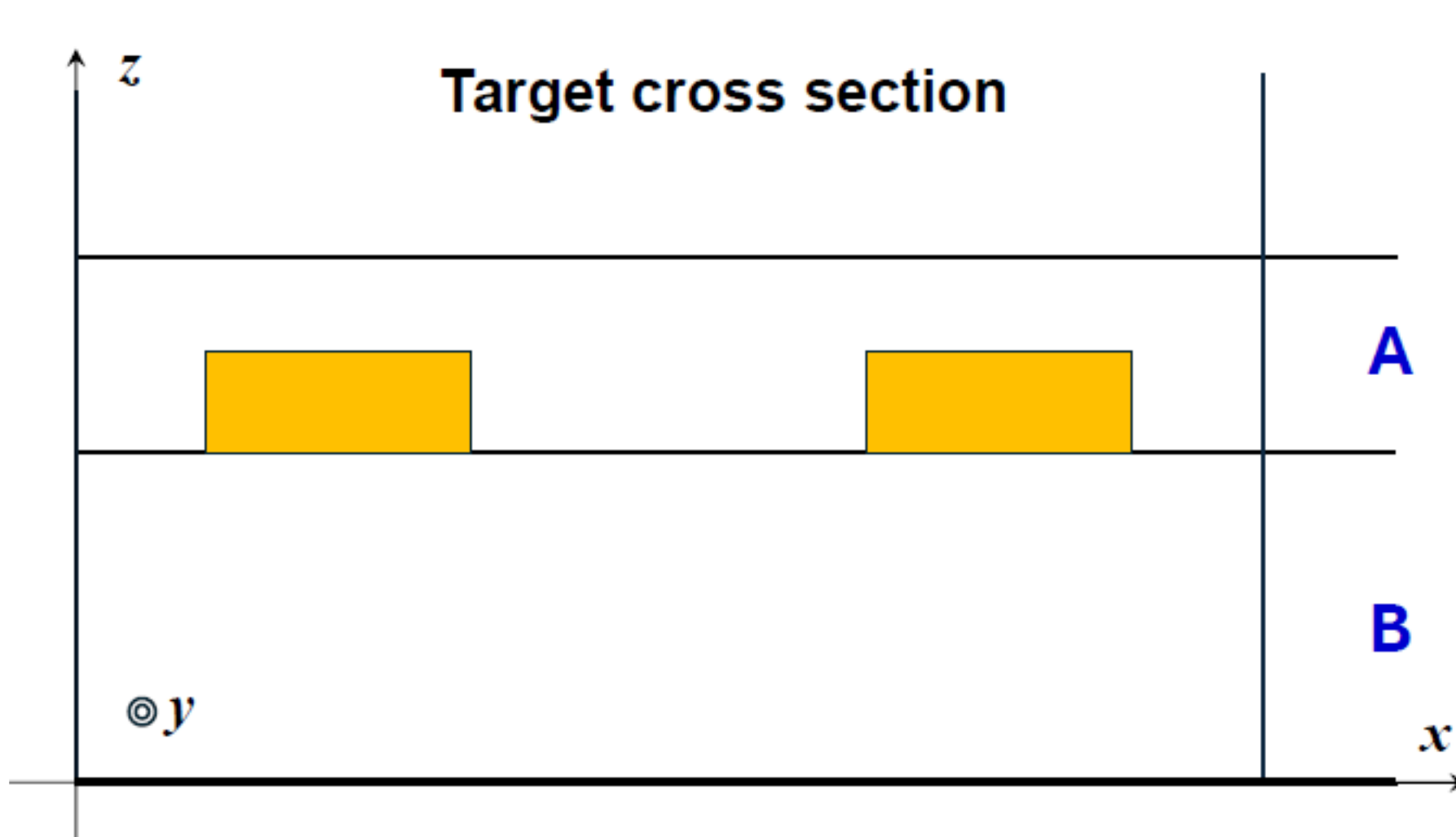

**Figure 1.** Cross section of the directed target laying in the [$x$, $y$]-plane (the dimensions are not proportional to the real sample). The polarization vector of the laser bean in these of the “horizontal” irradiation is in the $x$-direction. The substrate layer **B** is quartz, $SiO_2$, and its thickness is 1 mm. Nanorod antennas (102x30x30 nm in size) are prepared on top of the substrate by Electron Beam Lithography. A 170 nm PMMA layer **A** is prepared by spin coating, and it covers the whole substrate including the Au nanostructures.

In the presented experiment the 30x102x30 nm golden nanorod antennas have a lateral, $x$-directed, center to center distance of 500 nm, while in the vertical, $y$-directed, distance 500 nm. In the experiments presented here the target had one active layer with nanoantennas, while with the same technology targets with two active layers were also manufactured.

Recently the capability of plasmonic structures to use light to study deep subwavelength volumes was pointed out for design, simulation and experiments for different purposes also [9-17].

For future validation experiments nanocomposites containing aligned resonant nanorod antennas are also being prepared. Inspired by the works of Pérez-Juste et al [18], Van der Zande et al [19], and Wilson et al [20], the resonant nanorods will be dispersed into a PVA (poly(vinyl alcohol) matrix, where alignment can be enforced on the nanorods by exposing the thin nanocomposite films to simultaneous heating and stretching. As PVA can be considered a polar polymer [21], the gold nanorods, covered with a matching polar molecular capping, tend to align in a preferred direction upon heating and stretching the polymer. The alignment is also driven by the elongation of the PVA molecules [22, 23]. The advantage of the method is that the alignment of the nanorods can be selectively achieved in large surface areas (e.g. in the several $cm^2$ range) in different areas of the polymer nanocomposite film, thus deliberately creating surface regions with an alternating pattern of aligned/random nanorod orientations. As the alignment of the nanorod, e.g. the direction of its main axis, is influenced by the stretching direction, alternating areas with different orientations can also be achieved. Considering the price of PVA, and the up-scalability of the technology, this method can be an affordable alternative to mass produce targets with oriented nanorods, especially compared to direct writing methods such as electron-beam or ion-beam lithography techniques, where the yield (in terms of surface area) is much smaller as significantly higher fabrication times/costs [24].

## Evidence of directed proton acceleration

Recently at ELI-ALPS the above-mentioned directed targets were tested regarding the direction of their proton acceleration. For this purpose, Thomson Parabola (TP) detectors were used. Two such detectors were installed one forward (FWD) where the particles were detected for thin targets, and another one on the backward (BWD) side of the target where a BWD plume reached the TP. This was important when the target thickness was larger, plus a 1 mm thick quartz substrate, so that most of the material of the target crater was emitted into the BWD plume. In this configuration the directed nanorods interacted with the target material both during the impact of the laser beam and at the emission of the BWD plume formation.

The incoming laser beam, the outgoing plume, and the TP detectors are in the “horizontal” plane of the target chamber. The polarization, ***E***-vector is also in this plane. The electric field oscillates in 45º direction to the vertical, $[x,y]$-plane of incidence.

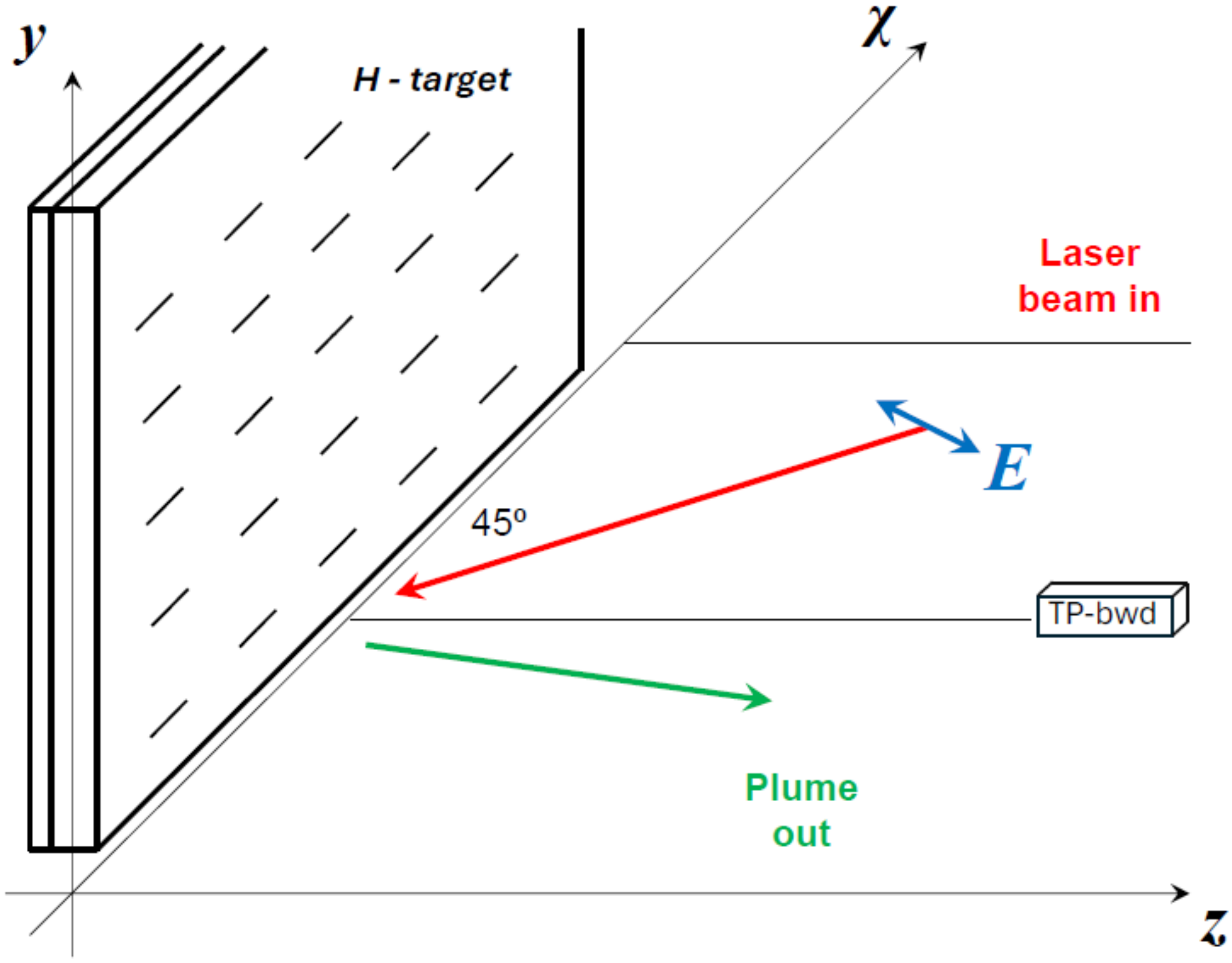

**Figure 2.** Configuration of the detection of protons accelerated by the directed nanorod antennas in a view from the top. The laser beam is indicated by the (red arrow). The polarization, $E$, of the laser beam (blue) is in the "horizontal" $[x,z]$-plane, orthogonal to the beam direction in the "horizontal" plane. The target is in the vertical, $[x,y]$-plane. The target is 170μm thick, so the plume is mostly emitted backwards orthogonally to the plane of the target (green arrow). The direction of the nanorod antennas is either in the "horizontal" $x$-direction (90°/270°) as indicated in the figure, or in the "vertical" $y$-direction.

As reported recently [25, 26] it is important to study the details of the observed proton (and other ion) acceleration. The acceleration by the nanorods is non-linear in the sense that the energy increase is not proportional with the increase of proton numbers, but at higher proton numbers the proton energies are increasing even more. Thus, the integrated energy increase is even larger than the integrated proton number increase.

In the experimental setup (Fig. 2), when the target nanorod antennas were "horizontal" thus close to parallel to the polarization of the laser beam, BWD directed TP shows much larger accelerated proton number density (~10 times) and at ~30-40 % higher energies as in the opposite case (Fig. 6). For "vertical" nanorod antennas when the laser beam polarization and the nanorod antennas were orthogonal then these antennas could not lead to any acceleration.

For the EPOCH simulation we chose a Calculation Box (CB) of 500 x 500 μm in the $[x, y]$-plane with periodic boundary condition, and in the z -direction we used, as in the experiment, only one layer of 170 μm thickness of Hydrogen (only Hydrogen for simplicity), Fig. **1**. In the middle of the bottom of the CB we placed a Gold nanorod antenna of cross section 30 x 30 nm and length of 102 nm pointing in the horizontal $x$ -direction for the "Horizontal" target. For the "Vertical" target it pointed in the $y$ -direction. The Hydrogen density was the same as in the PMMA resist matter, while the effect of C and O atoms were neglected for simplicity.

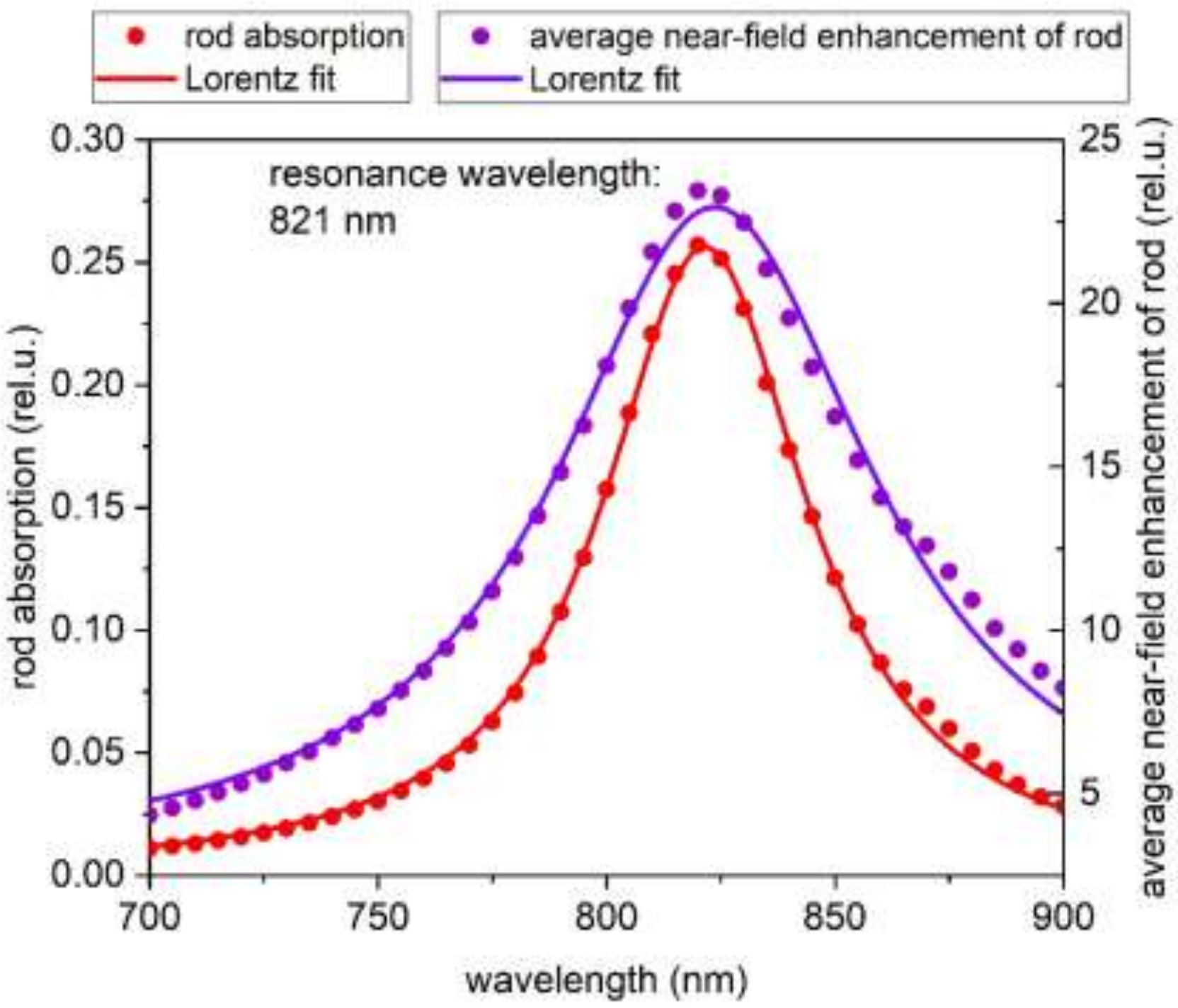

**Figure 3.** Resonance analysis of a 102 x 30 x 30 nm Gold nanorod antenna in PMMA resists under 45º laser irradiation, relative to the situation without nanorod antenna (rel. u.). In this configuration the vacuum resonance length is 821 nm, calculated in the COMSOL Multiphysics Simulation Software. In our experiment the laser irradiation has a vacuum

wavelength of 795 nm, and as we see the presently used antenna under the 45º laser irradiation is not the optimal resonance.

The resonance properties of these antennas were tested numerically by A. Szenes, M. Csete et al. See Fig. **3**. Due to the 45º irradiation (see Figs. **5** and **6**) the resonance length of the used nano-antenna is different from the vacuum wavelength of the laser beam.

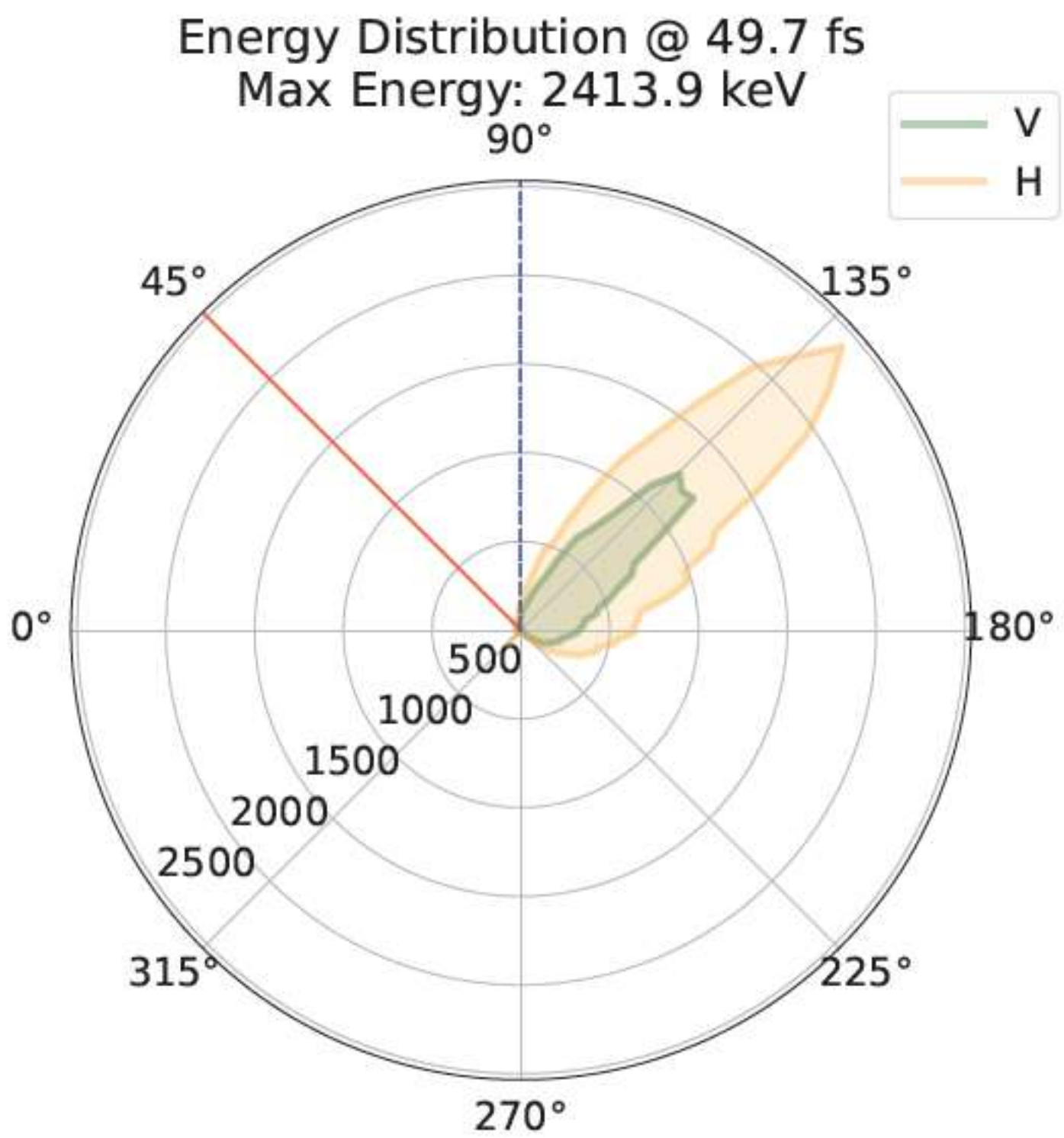

**Figure 4.** The calculated angular energy distribution of accelerated protons in the experimental configuration shown in Fig. **2**. The Incoming laser beam of intensity $4 \cdot 10^{21}$ W/cm$^2$ arrives from the 45º angle and due to the quartz substrate, the accelerated protons are emitted into the 135º degree direction. In case of horizontal nanorod antennas (pink contour line) pointing into the $\pm x$, (0-180º) direction the protons are accelerated more and enwidened also in the $\pm x$, direction. In case of the vertical nanorod antennas (green contour line) in the target, these antennas are orthogonal to the laser irradiation and to the laser polarization, thus these antennas have no effect on the proton acceleration, accelerated only to about half of the energy than in the horizontal case.

Due to the laser irradiation from the $z$ -direction we expect an enhanced proton emission into the plume in the reflected BWD $z$ -direction, while part of the original $x$ -component of the incoming laser beam still occurred. Due to the Horizontal $x$ -directed nanorods we expected an $x$ -directed dipole resonance along the nanorod with increasing amplitude with increasing irradiation time (Fig. **4**). For Vertical target with $y$ -directed nanorods this second effect vanishes as the polarization of the beam is orthogonal to the $y$ -direction.

The TP detector in the $z$ -direction unfortunately cannot measure the full energy of enhanced emission of protons in the $x$ -direction but the increased proton number due to the increased proton resonant emission, still makes the increased proton acceleration visible. On the other hand, the EPOCH simulation clearly shows the increased proton emission in the $x$ -direction, which is not visible in the present detector setup, Fig. **2**.

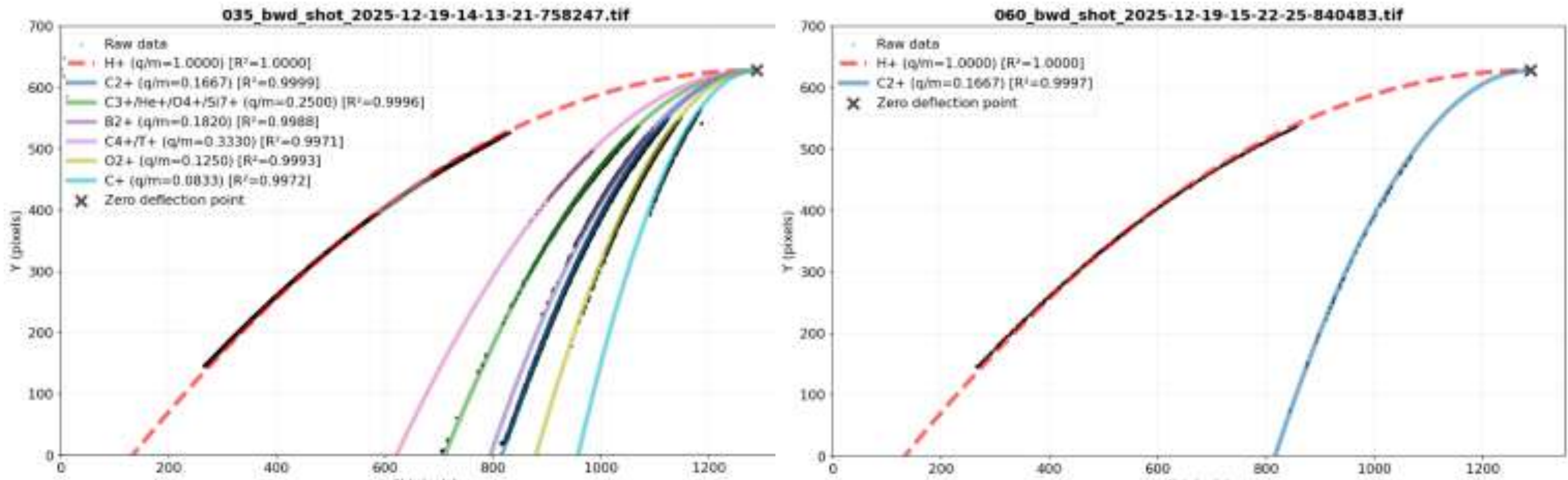

**Figure 5.** Thomson Parabola images of emitted protons (red) and positive ions arising from laser irradiation of directed nanorod targets with different nanorod directions. In the horizontal (left) case the nanorod antennas and the laser beam polarization were closer to parallel (45º), while in the vertical (right) case the two directions were orthogonal. The laser beam pulse energy was 30 mJ, the pulse duration was 120.5 fs, and the target thickness was 220 μm. The length of the nanorod antennas was 102 nm. The number and energy of accelerated protons, $2.4 \cdot 10^5$, is larger in the horizontal case when the laser beam polarization and the nanorod antennas are close to parallel. We see that the constituents of PMMA (C5, O2, H8)n, appear on the TP spectrum.

In Fig. **5** we can see that the Horizontal target nanorods are accelerating the protons and four other ions, while in the Vertical case the nanorods are orthogonal to the laser beam polarization. In Fig. **5**, however, the accelerated proton numbers and energies are not visible, and we see that protons are accelerated in both the Horizontal and Vertical cases. We plot the energy distribution of the protons for both cases in Fig, **6**.

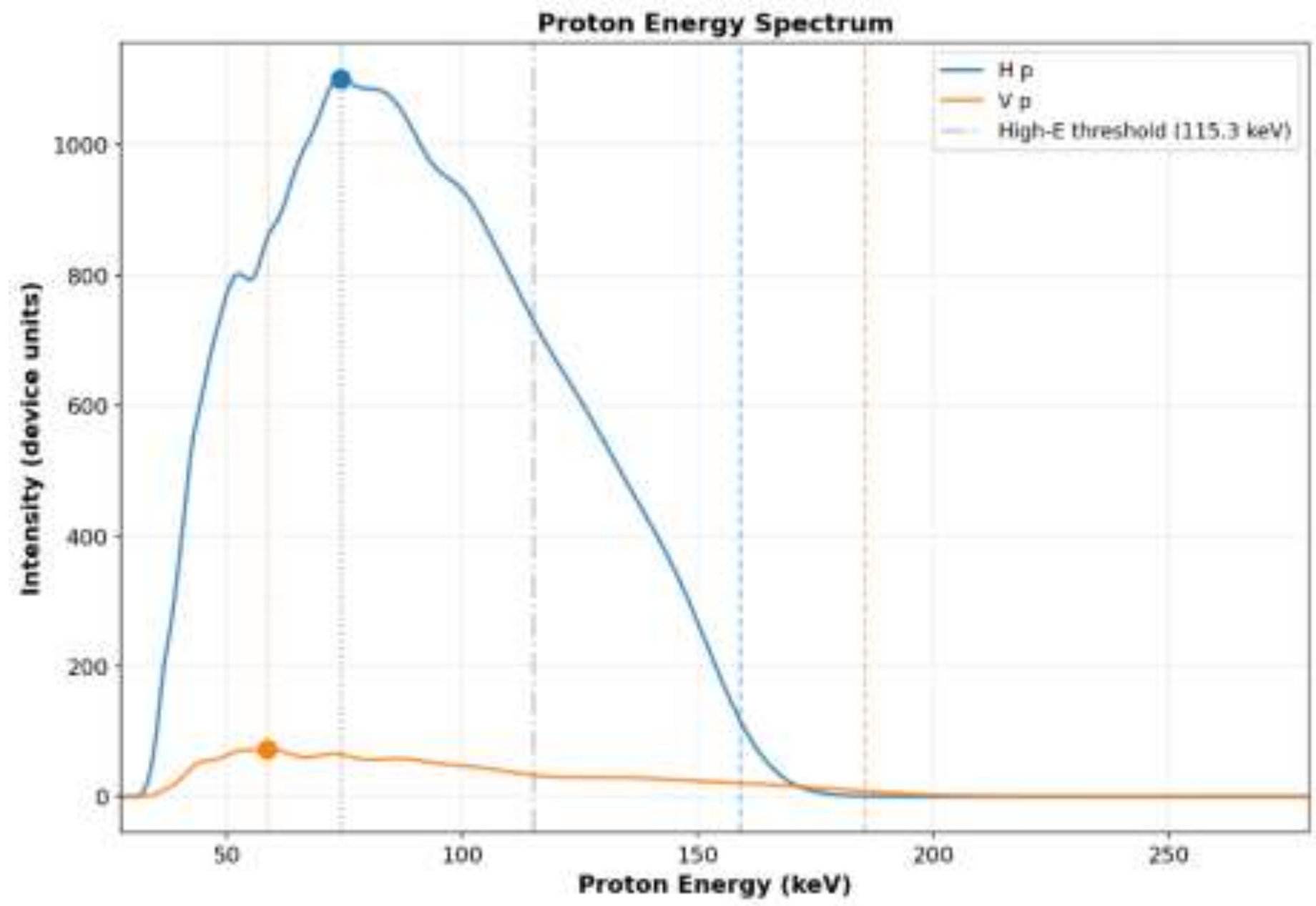

**Figure 6**. The energy spectrum of emitted protons with respect to the proton energy after irradiation of targets with Horizontal (H, blue line) and Vertical (V yellow line) nanorod targets, measured by the BKW TP detector in the $z$ -direction. The Horizontal target has nanorod antennas that are 45º from the polarization vector direction of the infalling laser light, while the Vertical target has nanorod antennas orthogonal to the polarization of the laser light. Partly parallel polarization leads to increased resonance and proton acceleration, both the energy and proton number are significantly larger in this case. The observed maximum proton energy is, 151 keV, the highest number of protons observed at 72.9 keV. The observed minimum proton energy is at 31.3 keV. The maximum number of accelerated protons in horizontal direction (closest to the polarization of the laser beam) is $2.4 \cdot 10^5$, more than an order of magnitude larger than in the vertical (orthogonal) direction.

The integrated total proton number detected the TP detector for the horizontal target was $8.07 \cdot 10^5$, about an order of magnitude more than the total detected proton number for the vertical target. This is a significant difference although the test configuration was not optimal as the polarization vector and the direction of the nanorod antennas were not fully parallel. Due to the 45º angle difference between the measured direction of protons and the direction of the polarization direction of the laser beam the energy of the protons (Fig. **6**) was less than the typical energy resulting in p+11B fusion [27].

As we can see in Fig. **5**, other heavier ions are also accelerated in the same direction as the protons, and this applies to fusion reaction products, like $\alpha$-particles also. For our fusion strategy this is important as we aim to avoid thermalization to minimize energy losses, and this is possible even at the last stage at the final product of the nuclear fusion reactions (Figs. **4** and **6**)!

## Considerations for fusion reactions

Our recent proton-boron experiments verified [27] that this way we can achieve fusion with relatively weak, 25 mJ, laser beam pulses at the ELI-ALPS European laser infrastructure in Szeged. This was possible even if we could not yet apply all the features we planned.

Our aim is to manufacture fusion targets for extremely rapid radiation dominated ignition of fusion with relatively smaller compression and keeping non-thermal processes minimal to reduce thermalization losses and maintain the ignition energy as much as possible until of ignition or even beyond. Measuring two-particle correlations may give insight to spatial and temporal distribution of fusion reactions [28].

This requires the following conditions:
(i) The target should be solid at room temperature, to avoid energy spent on cooling, and having less non-fusion material in the target. This way the energy spent on non-fusion materials will be minimized. E.g. for p+11B fusion we should use polymer targets with mostly Hydrogen and Boron content.
(ii) To achieve, in the most affordable way, rapid or simultaneous ignition of the whole fuel target we need two-sided laser irradiation timed to achieve constructive superposition in the middle of the collision. In other words, the time precision of collision should be one-fourth of the laser wave period or less, which is of the order of one femtosecond.
(iii) To achieve simultaneous ignition we need a laser irradiation pulse duration such that the laser beam penetrates the target thickness once and only once. This time depends on the speed of light in the target material, the total energy of one laser pulse and this gives the required target thickness. For small laser pulse energy (e.g. 25 mJ) in UDMA target material we may have about 20-40 μm thick target, while for 2.5 J laser pulse energy we could have about 2-4 mm target thickness. This makes the target manufacturing requirements easier! Important that the laser beam intensity should be the same and exceeding $10^{17} – 10^{18}$ W/cm$^2$.
(iv) From the previous two requirements a laser facility is required, which is tunable with fs precision or shorter and which is having a time contrast of similar duration. At present most laser facilities do not satisfy this requirement except ELI-ALPS, according to our information.
(v) To regulate and maximize the absorption of the laser beam energy we need a high gain antenna array system. This minimizes the required number of antennas and may avoid unwanted thermalization of laser energy, while keeping the accelerated proton energies as uniform and as parallel as possible.

The fusion fuel targets used up to now within the NAPLIFE project do not satisfy fully these requirements. We can go through the presently used and already manufactured targets and manufacturing technologies. Furthermore, we will discuss the possible other directions of fusion target development for radiation dominated ignition. In this direction the NAPLIFE (and recently the FUSENOW) projects are unique.

## Authors Contribution

Zs.M., J.B., At.B., G.H., participated primarily in target manufacturing, I.B., M.A, T.G., Á.I., M.K., K.O., P.R., Á.Sz., P.V., M.V., N.K, participated primarily in experimental activities, Al.B., M.Cs., M.G., J.H., I.P., A.Sz., D.V., K.Zs., L.Cs. participated primarily in theoretical

predictions and evaluations, T.S.B., participated primarily in project leadership, manuscript reading and editing.

## Acknowledgements

L. P. Csernai acknowledges support from Wigner Research Center for Physics, Budapest (2022-2.1.1-NL-2022-00002). T.S. Biró, M. Csete, N. Kroó, I. Papp, acknowledges support by the National Research, Development and Innovation Office (NKFIH) of Hungary. This work is funded in part by Tromsø Research Foundation through Trond Mohn Research Foundation's UiT initiative, FUSENOW (TMF2025UiT01). This work is supported in part by the Frankfurt Institute for Advanced Studies, Germany, the Hungarian Research Network, the Research Council of Norway, grant no. 255253, and the National Research, Development and Innovation Office of Hungary, for projects: Nanoplasmonic Laser Fusion Research Laboratory under project numbers (NKFIH-874-2/2020, 468-3/2021, 2022-2.1.1-NL-2022-00002), Optimized nanoplasmonics (K116362), and Ultrafast physical processes in atoms, molecules, nanostructures and biological systems (EFOP-3.6.2-16- 2017-00005).

## References

[1] A. H. Taub, Relativistic Rankine-Hugoniot Equations, *Phys, Rev*. **74**, 328 (1948).

[2] L.P. Csernai, Detonation on a time-like front for relativistic systems, *Zh. Eksp. Teor. Fiz.* **92**, 379- 386 (1987).

[3] L.P. Csernai and D.D. Strottman, Volume ignition via time-like detonation in pellet fusion, *Laser and Particle Beams*, **33**, 279-282 (2015).

[4] L.P. Csernai, N. Kroo, and I. Papp, Radiation dominated implosion with nano-plasmonics, *Laser and Particle Beams*, **36** (2), 171-178 (2018).

[5] E. Prodan, C. Radloff, N. J. Halas, and P. Nordlander, "A hybridization model for the plasmon response of complex nanostructures," Science. 302 (5644), 419–422 (2003).

[6] I. Papp and K. Zsukovszki, Particle simulation of various gold nanoantennas in laser-irradiated matter for fusion production (part of NAPLIFE Collaboration), *Eur. Phys. J. Spec. Top.* Jul. 2 (2025),

[7] István Papp, Larissa Bravina, Mária Csete, Archana Kumari, Igor N. Mishustin, Anton Motornenko, Péter Rácz, Leonid M. Satarov, Horst Stöcker, András Szenes, Dávid Vass, Tamás S. Biró, László P. Csernai, Norbert Kroó, on behalf of NAPLIFE Collaboration,

PIC simulations of laser-induced proton acceleration by resonant nanoantennas for fusion, *arXiv:*2306.13445v2 [physics.plasm-ph] https://doi.org/10.48550/arXiv.2402.2306.13445v2

[8] https://www.allresist.com/portfolio-item/e-beam-resist-ar-p-672-series/, Allresist. [Online]. [Accessed 20 08 2025].

[9] Sanjay Kumar, Sanju Tanwar, Sumit Kumar Sharma, Nanoantenna – A Review on Present and Future Perspective, *International Journal of Science, Engineering and Technology*, 4(1), 2395-4752 (2016).

[10] Tae-In Jeong, Dong Kyo Oh, San Kim, Jongkyoon Park, Yeseul Kim, Jungho Mun, Kyujung Kim, Soo Hoon Chew, Junsuk Rho, and Seungchul Kim, Deterministic nanoantenna array design for stable plasmon-enhanced harmonic generation, *Nanophotonics*; 12(3): 619–629 (2023). https://doi.org/10.1515/nanoph-2022-0365

[11] Aftab Ahmed, Yuanjie Pang, Ghazal Hajisalem, Reuven Gordon, Antenna Design for Directivity-Enhanced Raman Spectroscopy, *International Journal of Optics,* Volume **2012**, 729138, (2012). https://doi.org/10.1155/2012/729138

[12] Waleed Tariq Sethi, Olivier De Sagazan, Mohamed Himdi, Hamsakutty Vettikalladi and Saleh A. Alshebeili; Thermoelectric Sensor Coupled Yagi–Uda Nanoantenna for Infrared Detection, *Electronics*, 10, 527, (2021),

[13] Moshik Cohen, Reuven Shavit & Zeev Zalevsky, Observing Optical Plasmons on a Single Nanometer Scale; *Scientific Reports* **4**, 4096, (2014), https://doi.org/10.1038/srep04096

[[(21 February 2014); 1) Faculty of Engineering, Bar-Ilan University, Ramat-Gan 52900, Israel, 2) Department of Electrical and Computer Engineering, Ben-Gurion University of the Negev, Beer-Sheva 84105, Israel, 3) Bar-Ilan Institute for Nanotechnology &Advanced Materials, Ramat-Gan 52900, Israel.]]

[14] Wenxing Chen, Mykhailo Tymchenko, Prashanth, Gopalan, Xingchen Ye, Yaoting Wu, Mingliang Zhang, Christopher B Murray, Andrea Alu, Cherier R Kagan, “Large- Area Nanoimprinted colloidal Au Nanocrystal Based Nanoantenna for Ultrathin Polarizing Plasmonic metasurface”, *Nanoletter*: **15**(8), (2015), https://doi.org/10.3390/electronics10050527

[15] S. Kasani, K. Curtin, and N. Wu, A review of 2D and 3D plasmonic nanostructure array patterns: fabrication, light management and sensing applications, *Nanophotonics*, **8**(12), 2065−2089, (2019).

[16] Min Q., Pang Y., Collins D.J., Kuklev N.A., Gottselig K., Steuerman D.W., and Gordon R., Substrate-based platform for boosting the surface enhanced Raman of plasmonic nanoparticles, *Optics Express*, **19**(2), 1648-1655 (2011).

[17] Defeng Kong, Guoqiang Zhang, Yinren Shou, Shirui Xu, Zhusong Me, Zhengxuan Cao, Zhuo Pan, Pengjie Wang, Guijun Qi, Yao Lou, Zhiguo Ma, Haoyang Lan, Wenzhao Wang, Yunhui Li, Peter Rubovic, Martin Veselsky, Aldo Bonasera, Jiarui Zhao, Yixing Geng, Yanying Zhao, Changbo Fu, Wen Luo, Yugang Ma, Xueqing Yan and Wenjun Ma, High-

energy-density plasma in femtosecond-laser-irradiated nanowire array targets for nuclear reactions, *Matter Radiat. Extremes* **7**, 064403 (2022).
https://doi.org/10.1063/5.0120845

[18] J. Pérez-Juste, B. Rodríguez-González, P. Mulvaney, and L. M. Liz-Marzán, "Optical Control and Patterning of Gold-Nanorod–Poly(vinyl alcohol) Nanocomposite Films," *Adv Funct Mater*, **15**, no. 7, pp. 1065–1071, Jul. 2005, https://doi.org/10.1002/ADFM.200400591

[19] B. M. I. Van Der Zande, L. Pagès, R. A. M. Hikmet, and A. Van Blaaderen, "Optical Properties of Aligned Rod-Shaped Gold Particles Dispersed in Poly(vinyl alcohol) Films," *Journal of Physical Chemistry B*, **103**, no. 28, pp. 5761–5767, Jul. 1999, https://doi.org/10.1021/JP9847383

[20] O. Wilson, G. Wilson, and P. Mulvaney, "Laser writing in polarized silver nanorod films," *Advanced Materials*, **14**, no. 13–14, 2002, https://doi.org/10.1002/1521-4095(20020705)14:13/14<1000::AID-ADMA1000>3.0.CO;2-E

[21] S. Asano, "Polyvinyl Alcohol: A Comprehensive Overview," *Advanced Materials Science Research*, **7**, no. 5, pp. 199–200, Oct. 2024, https://doi.org/10.37532/aaasmr.2024.7(5).199-200

[22] D. Fornasiero and F. Grieser, "A linear dichroism study of colloidal silver in stretched polymer films," *Chem Phys Lett*, **139**, no. 1, pp. 103–108, Aug. 1987, https://doi.org/10.1016/0009-2614(87)80159-8

[23] J. Michl, E. W. Thulstrup, and J. H. Eggers, "Polarization spectra in stretched polymer sheets. III.1 Physical significance of the orientation factors and determination of π-π* transition moment directions in molecules of low symmetry," *J Phys Chem*, **74**(22) 3878–3884, 1970, https://doi.org/10.1021/j100716a006

[24] B. Abasahl, C. Santschi, T. V. Raziman, and O. J. F. Martin, "Fabrication of plasmonic structures with well-controlled nanometric features: a comparison between lift-off and ion beam etching," *Nanotechnology*, vol. 32, no. 47, p. 475202, Aug. 2021, https://doi.org/10.1088/1361-6528/AC1A93

[25] Imene Benabdelghani, Márk Aladi, Péter Rácz, Gergő Hegedüs, Ádám Inger, László P. Csernai, Tamás Sándor Biró, Norbert Kroó, Miklós Ákos Kedves, Károly Osvay, Parvin Varmazyar, Tibor Gilinger, Attila Bonyár, Nóra Tarpataki, Zsuzsanna Márton (NAPLIFE collaboration), Experimental investigations of laser-driven ion acceleration and fusion reaction enhanced by Plasmonic Nanostructured Targets, invited talk, XIV International Conference on New Frontiers in Physics, 17-31 July 2025, OAC, Kolymbari, Crete, Greece

[26] Laszlo P. Csernai (for the NAPLIFE & FUSENOW Collaborations), Laser Induced p+11B Fusion by Resonant Nanoplasmonic Antennas, invited talk, XIV International Conference on New Frontiers in Physics, 17-31 July 2025, OAC, Kolymbari, Crete, Greece

[27] N. Kroó, L.P. Csernai, I. Papp, M.A. Kedves, M. Aladi, A. Bonyár, M. Szalóki, K. Osvay, P. Varmazyar, T.S. Biró (for the **NAPLIFE Collaboration**), Indication of p + 11B Reaction in Laser Induced Nanofusion Experiment, *Scientific Reports* (Nature) **14**, 30087 (2024).,

[28] L.P. Csernai, T. Csörgő, I. Papp, M. Csete, J.P. Hansen, A. Szenes, K. Tamosiunas, D. Vass, T.S. Bíró, M. Veres and N. Kroó (part of NAPLIFE and FUSENOW Collaborations), Radiation pattern and source size of particles in nanoplasmonic fusion, *International Journal of Modern Physics* E (2026) 2642001 (17 pages) in press
https://www.worldscientific.com/doi/10.1142/S0218301326420012